\documentclass[aps,prl,singlecolumn,showkeys,showpacs,superscriptaddress,groupedaddress]{revtex4}  % for review and submission
\pdfoutput=1
\usepackage{graphicx}  % needed for figures
\usepackage{dcolumn}   % needed for some tables
\usepackage{bm}        % for math
\usepackage{amssymb}   % for math
\usepackage{setspace}
\usepackage{amsmath}
\usepackage{color}
\begin{document}
\begin{spacing}{1.5}
% The following information is for internal review, please remove them for submission
\widetext
\leftline{Primary authors: W. Chen}
\leftline{Current version 3.0}
\rightline{Email: chenw@swip.ac.cn}
%\leftline{To be submitted to Physical Review Letters }
%\leftline{Comment to {\tt d0-run2eb-nnn@fnal.gov} by xxx, yyy}
%\centerline{\em D\O\ INTERNAL DOCUMENT -- NOT FOR PUBLIC DISTRIBUTION}

% the following line is for submission, including submission to the arXiv!!
%\hspace{5.2in} \mbox{Fermilab-Pub-04/xxx-E}

%\title{Experimental Verification of AITG Modes in a Tokamak Plasma with Weak Magnetic Shears}
%\title{Finding and Understanding of AITG Activities in Weak Magnetic Shear Plasmas}
\title{Alfv{\'e}nic Ion Temperature Gradient Activities in a Weak Magnetic Shear Plasma}

% remove these 3 lines before journal submittal.
%\centerline{author list dated 9 February 2012}
% end removal before journal submittal
%

\affiliation{Southwestern Institute of Physics, P.O. Box 432 Chengdu 610041, China}
\affiliation{Institute for Fusion Theory and Simulation, ZJU, Hangzhou 310027, China}
\affiliation{School of Physics and Optoelectronic, DUT, Dalian 116024, China}
\author{W. Chen$^1$, R.R. Ma$^1$, Y.Y. Li$^2$, Z.B. Shi$^1$, H.R. Du$^3$, M. Jiang$^1$, L.M. Yu$^1$, B.S. Yuan$^1$, Y.G. Li$^1$, Z.C. Yang$^1$, P. W. Shi$^1$, X.T. Ding$^1$, J.Q. Dong$^1$, Yi. Liu$^1$, M. Xu$^1$, Y.H. Xu$^1$, Q.W. Yang$^1$, and X.R. Duan$^1$}
\noaffiliation
\vskip 0.25cm
       % D0 authors (remove the first 3 lines
                             % of this file prior to submission, they
                             % contain a time stamp for the authorlist)
                             % (includes institutions and visitors)
\date{\today}

\begin{abstract}
Abstract--We report the first experimental evidence of Alfv{\'e}nic ion temperature gradient (AITG) modes in HL-2A Ohmic plasmas. A group of oscillations with $f=15-40$ kHz and $n=3-6$ is detected by various diagnostics in high-density Ohmic regimes. They appear in the plasmas with peaked density profiles and weak magnetic shear, which indicates that corresponding instabilities are excited by pressure gradients. The time trace of the fluctuation spectrogram can be either a frequency staircase, with different modes excited at different times or multiple modes may simultaneously coexist. Theoretical analyses by the extended generalized fishbone-like dispersion relation (GFLDR-E) reveal that mode frequencies scale with ion diamagnetic drift frequency and $\eta_i$, and they lie in KBM-AITG-BAE frequency ranges. AITG modes are most unstable when the magnetic shear is small in low pressure gradient regions. Numerical solutions of the AITG/KBM equation also illuminate why AITG modes can be unstable for weak shear and low pressure gradients. It is worth emphasizing that these instabilities may be linked to the internal transport barrier (ITB) and H-mode pedestal physics for weak magnetic shear.
\end{abstract}

\email{chenw@swip.ac.cn}

\maketitle

Kinetic Alfv{\'e}n and pressure gradient driven instabilities are very common in magnetized plasmas both in space and laboratory\cite{gvlc}\cite{czrmp}\cite{wu12}. In present-day fusion and future burning plasmas, they are easily excited by energetic particles (EPs) and/or pressure gradients. They can not only cause the loss and redistribution of EPs but also affect plasma confinement and transport\cite{lauberpr}\cite{nsk}. The physics associated with them is an intriguing but complex area of research. For weak magnetic shear ($s=(r/q)(dq/dr)\sim0$) and low pressure gradients ($\alpha=-R_0q^2d\beta/dr<1$; with $\beta$ the ratio of kinetic to magnetic pressures.), the stability and effect of them, such as Alfv{\'e}nic ion temperature gradient (AITG) mode\cite{zoncapop99}\cite{dongpop99}/kinetic ballooning mode (KBM)\cite{chengczpof}, have not been hitherto unrecognized. At weak magnetic shear, the first pressure gradient threshold becomes very small or vanishes and the AITG/KBM spectrum is unstable in the very low pressure gradient region\cite{hmprl}\cite{hmpop}. For equilibria with reverse shear where $q_{min}$ is off axis and $\alpha_{max}$ near $q_{min}$, there exists an unstable low-n global branch of AITG and trapped electron dynamics can further destabilize it\cite{gvprl}. The AITG/KBM modes, on the one hand,  can cause cross-field plasma transport that set an upper limit on the plasma beta; on the other hand, this electromagnetic turbulence could be a paradigm which can bridge electron and ion transport channels via finite $\beta$-effects. For the case of weak magnetic shears and low pressure gradients, so far, no clear experimental evidences supported this theoretical understanding, based on analytical and numerical simulation results. This paper presents a direct experimental evidence of AITG existence and corresponding physics mechanisms of mode excitation in tokamak plasmas.

\begin{figure}[!htbp]
\centering
\includegraphics[scale=0.8]{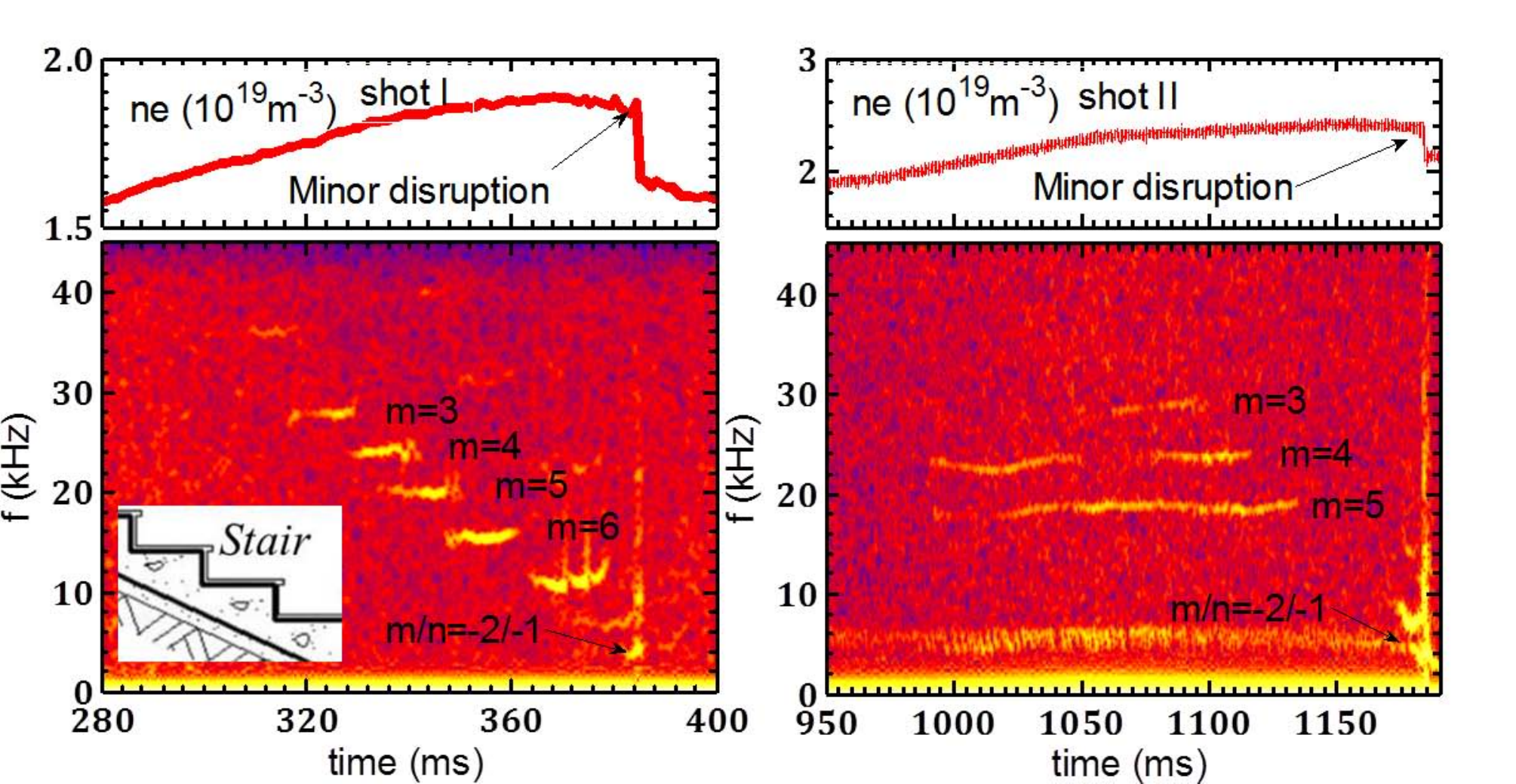}
\caption{\label{fig1}Typical discharges with AITG activities on HL-2A. 2D patterns are spectrograms of ECEI (left) and soft X-ray signal. \emph{Left} col.(shot I, Bt=1.31T, Ip=150kA) with the frequency staircase; \emph{Right} col.(shot II, Bt=1.35T, Ip=150kA) with the multi-mode coexistence.}
\end{figure}

The experiments discussed here are performed in deuterium plasmas with plasma current $I_p \simeq 150-170kA$, toroidal field $B_t \simeq 1.32-1.40T$, and an edge safety factor $q_a \simeq 4.2-4.8$ on HL-2A, which has the major/minor radius $R_0/a = 1.65 m/0.4 m$. The HL-2A plasma is almost a circular cross-section equilibrium although it corresponds to a divertor configuration in all during the discharges. The electron density was detected using a multi-channel HCOOH laser interferometer\cite{liyg}. The polodial mode number $m$ is measured using the electron cyclotron emission imaging (ECEI) signals by the spatial two-point correlation method\cite{jiangm}. The safety factor profile is obtained by the current filament code combined with far infrared (FIR) polarimetry data. This electromagnetic instability is observed only in high core density Ohmic plasmas, especially with peaked density profiles in limiter or divertor configuration. The phenomenon is perfectly reproducible. Mode features, including its frequency, mode-number and propagation direction, can be observed by ECEI, soft X-ray and microwave interference signals, respectively\cite{shipw}.
Figure 1 shows two typical experimental results during the plasma density ramp-up. Many coherent MHD fluctuations are visible around $f=15-40 kHz$ at $t=320-360 ms$ for shot I and $t=1000-1150 ms$ for shot II. These fluctuations do not appear on Mirnov signals. The poloidal mode number is obtained by the relation $m = Lk_\theta  /2\pi$, where $k_\theta$ is the poloidal wave vector and $L$ is the distance between two poloidal ECEI signals. These coherent modes have typical poloidal mode number m=3-6 and wave vector $k_\theta=0.2-0.6 cm^{-1}$, and propagate poloidally in the ion diamagnetic drift direction, e.g. $m>0$. Occasionally, the nonlinear behavior of modes, which is shown in Figure 2, can be observed during the frequency staircase. It is found that the mode frequency has a characteristic with the chirping-up. It suggests that the mode is an Alfv{\'e}nic instability, i.e., the mode is electromagnetic but electrostatic. To identify instabilities causing these fluctuations, we need to determine local plasma parameters. In the high core density Ohmic regime, we assume $T_i=T_e$ and $n_i=n_e$. Figure 3 gives electron temperature, density and safety factor profiles during the coherent modes at two different times ($t1$ and $t2$). Figure 3 shows that magnetic shear is weak during MHD activities and $q_{min}\sim 1$. Further, the soft X-ray array measurements also indicate there are q=1 rational surfaces. According to the q-profile, radial mode localization and poloidal mode number, we determine the toroidal mode-number $n=m/q\sim m$. At q=1 surface, $T_i\sim 0.6keV$, $\rho_{q=1}\simeq 0.22$, $-R_0 \bigtriangledown ln n_i\equiv 1/\varepsilon_n \simeq 6.0$, $\eta_i \equiv \nabla \ln T_i/\nabla \ln n_i \simeq 1.5$, $\beta_i\sim 0.3\%$, $\alpha\sim 0.1$ and $|s|\ll 1$. As anticipated earlier, the observed electromagnetic instabilities are unstable in regions of weak shear and low pressure gradients.

\begin{figure}[!htbp]
\centering
\includegraphics[scale=0.7]{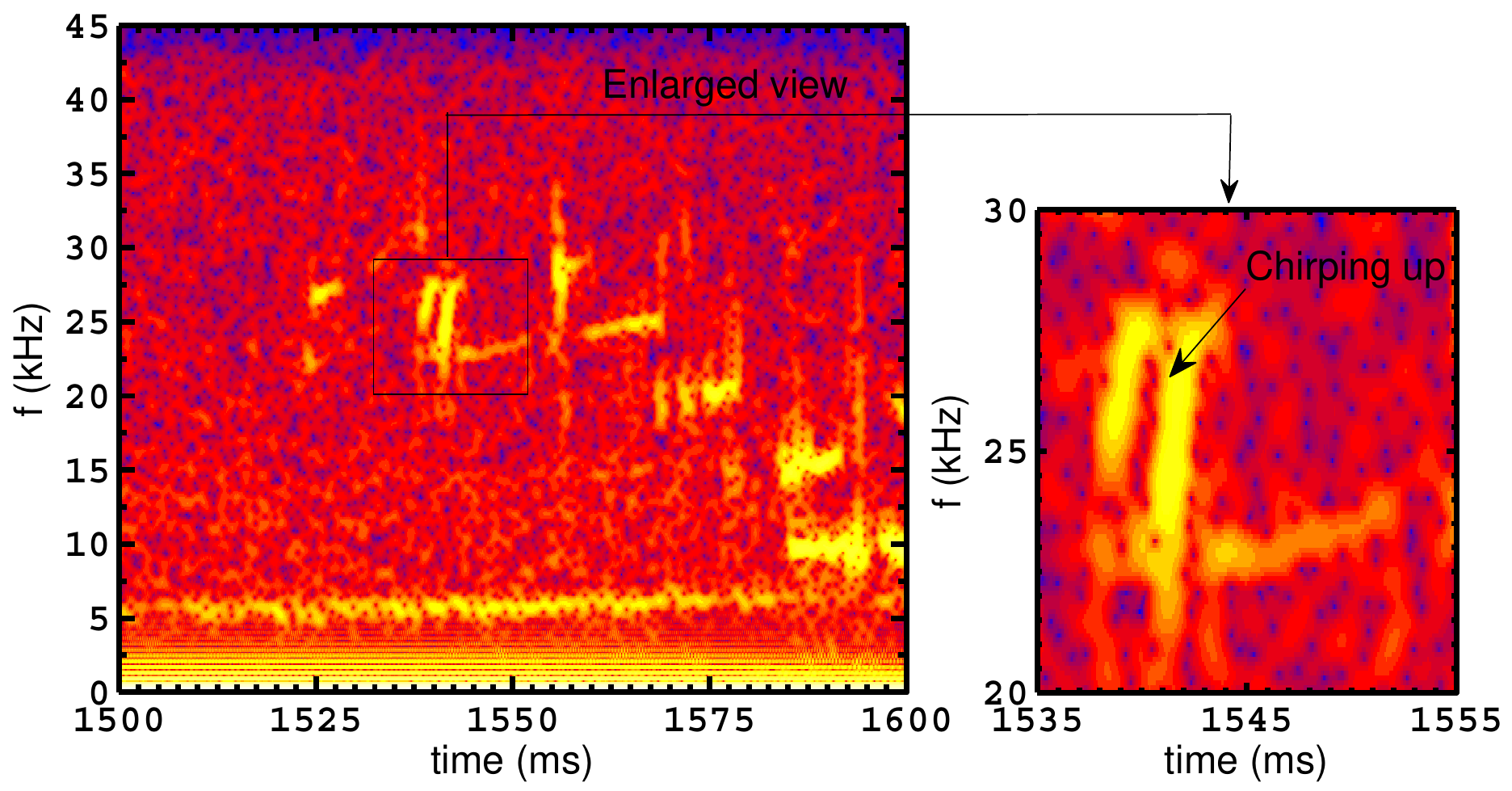}
\caption{\label{fig1} Nonlinear evolution of AITG activities during the frequency staircase.}
\end{figure}

The coherent modes occur in Ohmic plasmas with peaked density profiles and without any EPs, and they sometime have the behaviors of the frequency staircase. These observations suggest that they are possibly driven by pressure gradients and there is a threshold for mode excitation. To verify and understand them, we adopt the extended general fishbone-like dispersion relation (GFLDR-E) and AITG/KBM equation in the absence of EPs, respectively. Both these analytical theories assume the local $s-\alpha$ model equilibrium for shifted circular magnetic surfaces\cite{cht78}.

\begin{figure}[!htbp]
\centering
\includegraphics[scale=0.8]{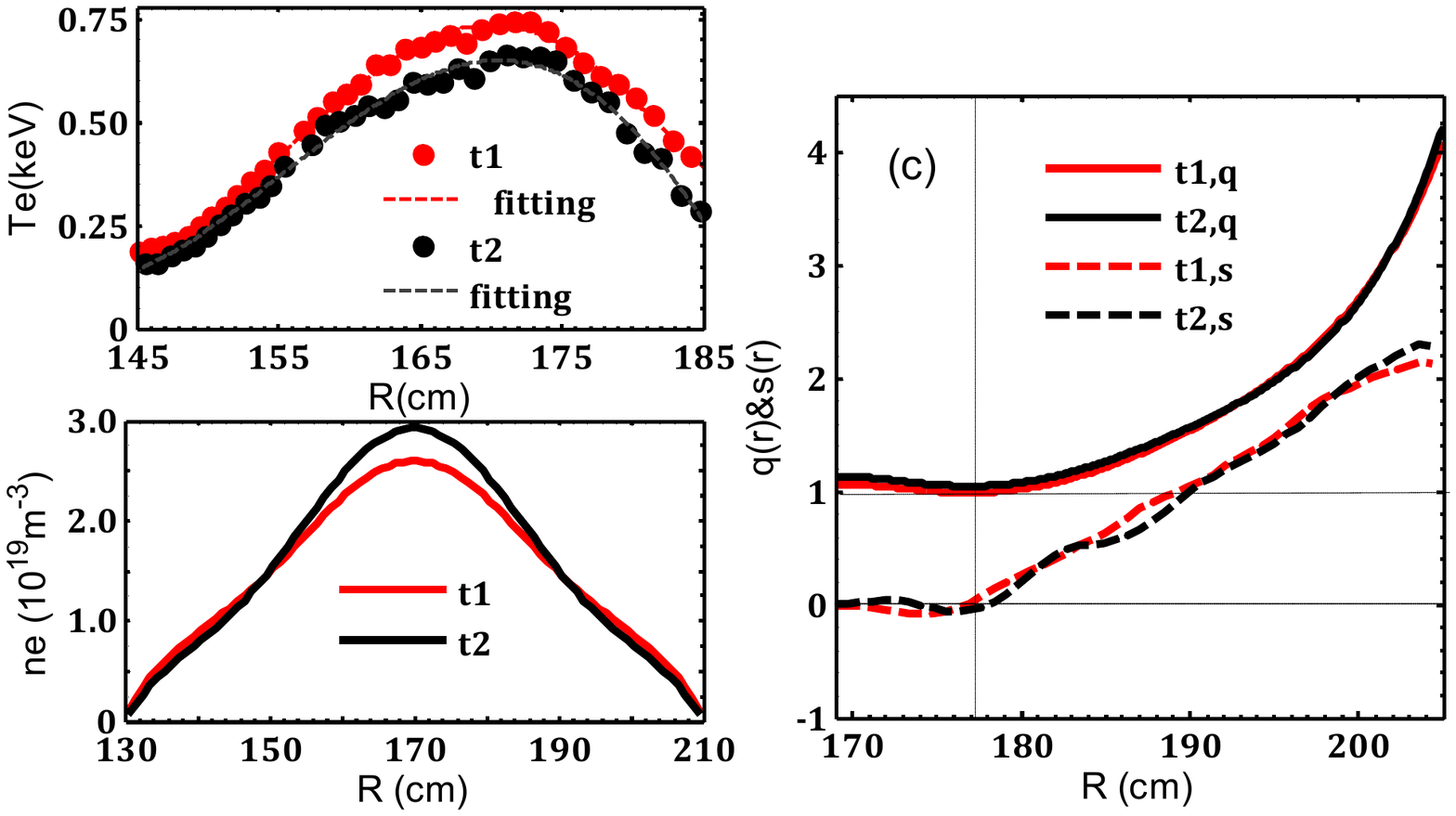}
\caption{\label{fig1} Electron temperature (a), density (b) and safety factor (c) profiles during observation of coherent modes.}
\end{figure}

\begin{table}[!hbp]
\small
\centering
\caption{Comparison between measured frequency and real frequency from the GFLDR-E. }
\begin{tabular}{ccccc}
%\begin{tabular}{|p{.2\textwidth}| p{.2\textwidth}| p{.2\textwidth}| p{.2\textwidth}|}
\hline
\multicolumn{1}{c}{$m/n$}&\multicolumn{1}{c}{$f_{Lab}=f_{MHD}-nf_{v \phi}$}& \multicolumn{1}{c}{$f_{v \phi} $}&\multicolumn{1}{c}{$f_{MHD}$}& \multicolumn{1}{c}{$\omega_r/2\pi$}\\
\hline
\multicolumn{1}{c}{$3/3$}&\multicolumn{1}{c}{$29$}& \multicolumn{1}{c}{$5$}&\multicolumn{1}{c}{$44$}& \multicolumn{1}{c}{$42$}\\
\hline
\multicolumn{1}{c}{$4/4$}&\multicolumn{1}{c}{$24$}& \multicolumn{1}{c}{$5$}&\multicolumn{1}{c}{$44$}& \multicolumn{1}{c}{$43$}\\
\hline
\end{tabular}

\small \textcolor[rgb]{0.00,0.00,0.00}{Note: $f_{v\phi}$ is the toroidal rotation frequency at q=1 surface.}
\end{table}

The GFLDR-E, which is derived by Zonca$\&$Chen and includes various kinetic effects\cite{zcsppcf96}\cite{zcpop14}, such as  finite Larmor radii (FLR) and finite orbit widths (FOW), etc., can be written as follows:
\begin{equation}
\begin{aligned}
- 2\sqrt Q \frac{{\Gamma (\frac{3}{4} - \frac{{\Lambda ^2 }}{{4Q}})}}{{\Gamma (\frac{1}{4} - \frac{{\Lambda ^2 }}{{4Q}})}} = \delta W_f
\end{aligned}
\end{equation}

where $\Lambda ^2  = \frac{{\omega ^2 }}{{\omega _A^2 }}(1 - \frac{{\omega _{ * pi} }}{\omega }) + q^2 \frac{{\omega \omega _{ti} }}{{\omega _A^2 }}[(1 - \frac{{\omega _{ * ni} }}{\omega })F - \frac{{\omega _{ * Ti} }}{\omega }G - \frac{{N^{\rm{2}} }}{D}]$, $Q^2 (\omega ) = s^2 k_\theta ^2 \rho _{i}^2 \frac{{\omega ^2 }}{{\omega _A^2 }}[\frac{3}{4}(1 - \frac{{\omega _{ * pi} }}{\omega } - \frac{{\omega _{ * Ti} }}{\omega }) + q^2 \frac{{\omega _{ti} }}{\omega }S(\omega ) + \frac{{(\Lambda \omega _A /\omega )^4 }}{{1/\tau  + (\omega _{ * ni} /\omega )}}]$, $S(\omega ) = \frac{{q^2 }}{2}(\frac{{\omega _{ti} }}{\omega })^2 [(1 - \frac{{\omega _{ * ni} }}{\omega })(L - 2L_{1/2}  - \frac{{2N}}{D}(H - 2H_{1/2} ) + \frac{{N^2 }}{{D^2 }}(F - 2F_{1/2} ))- \frac{{\omega _{ * Ti} }}{\omega }(M - 2M_{1/2}  - \frac{{2N}}{D}(I - 2I_{1/2} ) + \frac{{N^2 }}{{D^2 }}(G - 2G_{1/2} ))] + \frac{{q^2 }}{{D_{1/2} }}(\frac{{\omega _{ti} }}{\omega })^2 [(1 - \frac{{\omega _{ * ni} }}{\omega })(F_{1/2}  - F) - \frac{{\omega _{ * Ti} }}{\omega }(G_{1/2}  - G) - \frac{N}{D}(N_{1/2}  - N)]^2 + (1 - \frac{{\omega _{ * ni} }}{\omega })(T - \frac{{2N}}{D}V + \frac{{N^2 }}{{D^2 }}Z) - \frac{{\omega _{ * Ti} }}{\omega }(U - \frac{{2N}}{D}W + \frac{{N^2 }}{{D^2 }}(V - Z/2))$, $\omega _{ * pi}  = \omega _{ * ni}  + \omega _{ * Ti}  = (T_i /eB)k_\theta  (\nabla \ln n_i )(1 + \eta_i )$, $\omega _{ * ni}  = (T_i c/eB)(\vec k \times \vec b) \cdot \nabla \ln n_i $, $\omega _{ * Ti}  = (T_i c/eB)(\vec k \times \vec b) \cdot \nabla \ln T_i $, $\eta_i  = \nabla \ln T_i /\nabla \ln n_i $, $\tau=T_e/T_i$, $\omega_{ti}$ is the ion transit frequency, all functions, such as $F$ and $G$, are given by Ref.\cite{zcsppcf96}, and all symbols are standard. By a trial function method, $\delta W_f (\theta _k ) \simeq \frac{\pi }{{4\left| s \right|}}[s^2  - \frac{3}{2}\alpha ^2 \left| s \right| + \frac{9}{{32}}\alpha ^4  - \frac{5}{2}\alpha e^{ - 1/\left| s \right|} \cos (\theta _k ) - \frac{{5\alpha ^2 }}{{2\left| s \right|}}e^{ - 2/\left| s \right|} \cos (2\theta _k )]$ for $\alpha \sim\mid s\mid \sim O(1)$\cite{mzcpop15}. $\theta_k(ndq/dr)^{-1}k_r$ is the normalized radial wave vector, and $\alpha  = q^2 \beta _i [(1 + \eta _i ) + \tau (1 + \eta _e )]R_0 /L_n$ is the pressure gradient. The GFLDR-E shows that the shear Alfv{\'e}n-acoustic continuum structure can be modified by diamagnetic drift effects with strong density and temperature gradients, so that there is a transition from the beta-induced Alfv{\'e}n eigenmode (BAE) branch to the pure pressure-gradient driven and small scale kinetic ballooning mode (KBM) branch. Furthermore, finite $\nabla T_i$-effects give rise to another potentially unstable branch of Alfv{\'e}nic fluctuations, namely, the AITG mode due to wave-particle interactions with thermal ions via geodesci curvature coupling that is most unstable when the condition $\Omega_{*pi}\equiv\omega_{*pi}/\omega_{ti}\sim \sqrt{7/4+\tau}q$ is fulfilled. It can be understood as a branch connecting KBM (diamagnetic effects $\Omega_{*pi}\gg\sqrt{7/4+\tau}q$) and BAE (ion compression effects $\Omega_{*pi}\ll\sqrt{7/4+\tau}q$). Meanwhile, the AITG/KBM equation without trapped electron effects derived by Hirose\cite{hmpop} can be given below,
\begin{equation}
\begin{aligned}
\begin{array}{l}
 \frac{d}{{d\theta }}\{ [1 + (s\theta  - \alpha \sin \theta )^2 ]\frac{{d\phi }}{{d\theta }}\}  \\
  + \frac{\tau }{{2(1 + \tau )}}\frac{\alpha }{{\varepsilon _n (1 + \eta )}}\{ (\Omega  - 1)[\Omega  - f(\theta )] + \eta _e f(\theta ) \\
  - \frac{{(\Omega  - 1)^2 }}{{1 + \tau [1 - I_i (\theta )]}}\} \phi  = 0 \\
 \end{array}
\end{aligned}
\end{equation}

where $f(\theta ) = 2\varepsilon _n [\cos \theta  + (s\theta  - \alpha \sin \theta )\sin \theta ]$, $\Omega\equiv \omega/ \omega_{*e}$, $\eta  = (\eta _i  + \tau \eta _e )/(1 + \tau )$, $I_i (\theta ) = \frac{2}{{\sqrt \pi  }}\int_0^\infty  {xdx} \int_{ - \infty }^\infty  {ydy\frac{{\tau \Omega  + 1 + \eta _i (x^2  + y^2  - 3/2)}}{{\tau \Omega  + f(\theta )(x^2 /2 + y^2 )}}} J_0^2 [\sqrt 2 \xi (\theta )x]e^{ - x^2  - y^2 }$, and $\xi (\theta ) = k_\theta  \rho _i \sqrt {1 + (s\theta  - \alpha \sin \theta )^2 }$, and all symbols are also standard and given by Ref.\cite{hmpop}.

\begin{figure}[!htbp]
\centering
\includegraphics[scale=0.8]{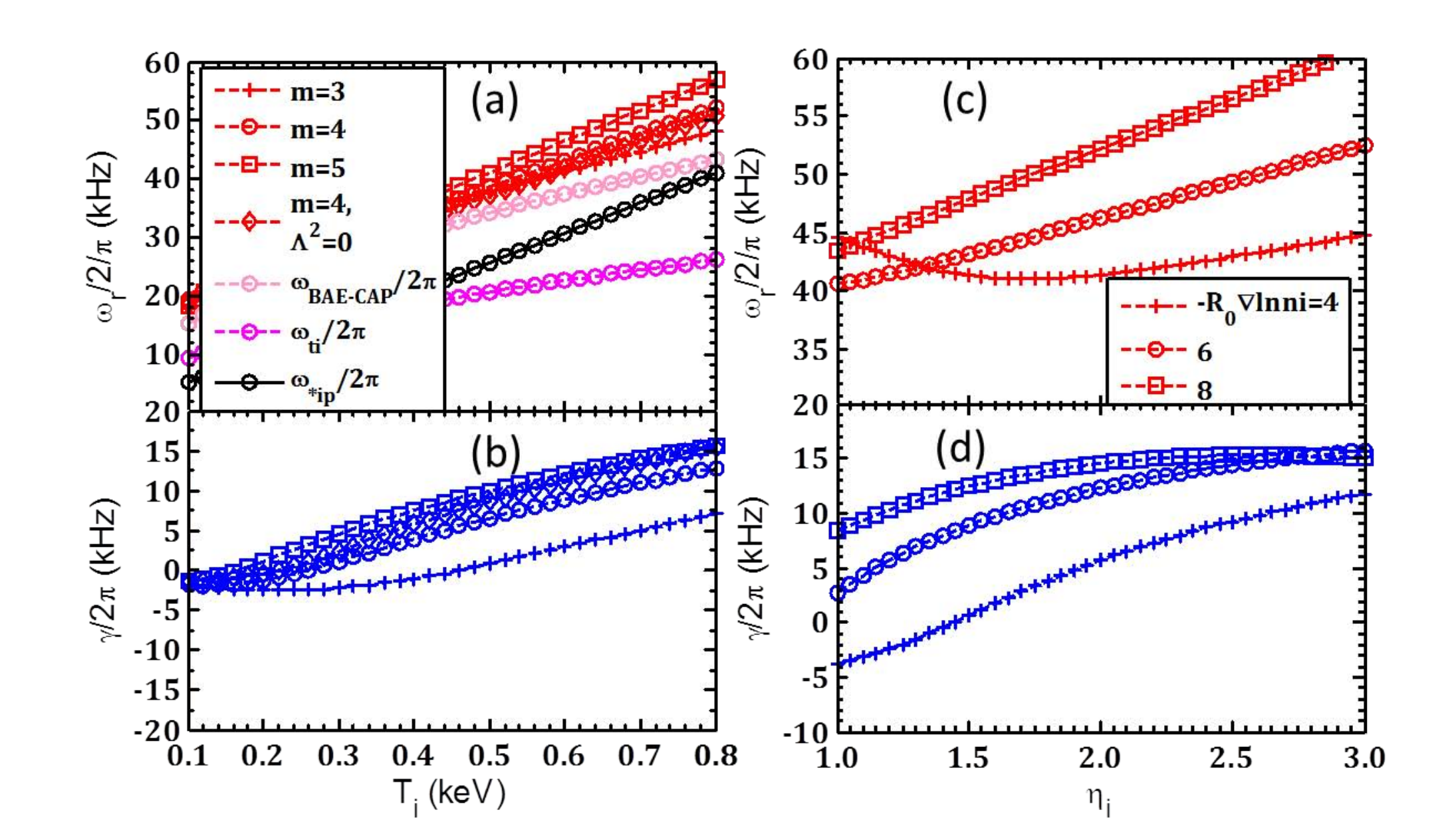}
\caption{\label{fig1} Solutions of the GFLDR-E according to parameters at q=1 surface, $s=-0.05$, $k_{\theta}\rho_i=0.1$ and $\theta_k=0$. Real frequecy (a) and growth rate (b) vs $T_i$; Real frequecy (c) and growth rate (d) of m/n=4/4 mode vs $\eta_i$. $\Lambda^2=0$ denotes the marginal stability, and all other situations $\delta W_f$ is from the approximate expression given in Ref\cite{mzcpop15}.}
\end{figure}

\begin{figure}[!htbp]
\centering
\includegraphics[scale=0.8]{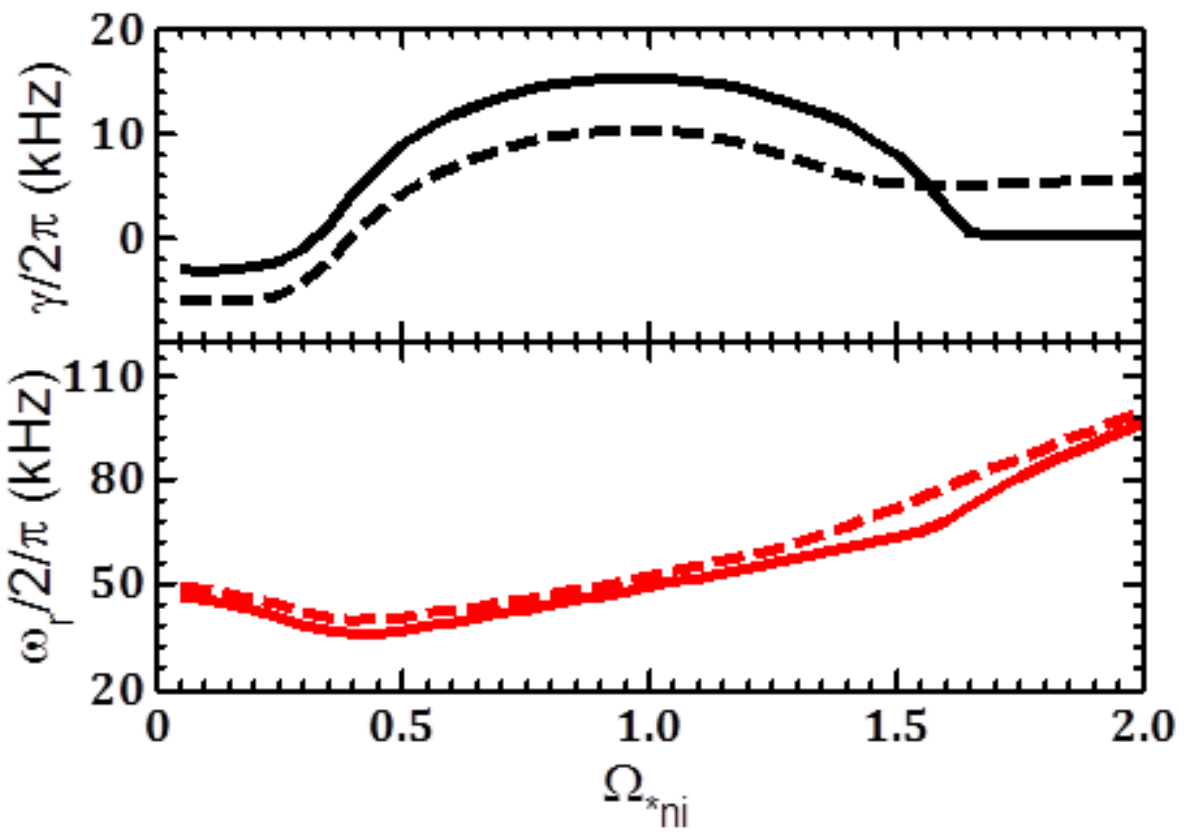}
\caption{\label{fig1} Diamagnetic effect ($\Omega_{*ni}\equiv \omega_{*ni}/ \omega_{ti}$) from density gradient on the stabity of coherent modes. Parameters from the q=1 surface, $s=-0.05$, $k_{\theta}\rho_i=0.1$ and $\theta_k=0$. Solid line: $\Lambda^2=0$; Dash line: GFLDR-E, $\delta W_f\neq 0$. Observed value $\Omega_{*ni}\sim 0.54$.}
\end{figure}

\begin{figure}[!htbp]
\centering
\includegraphics[scale=0.9]{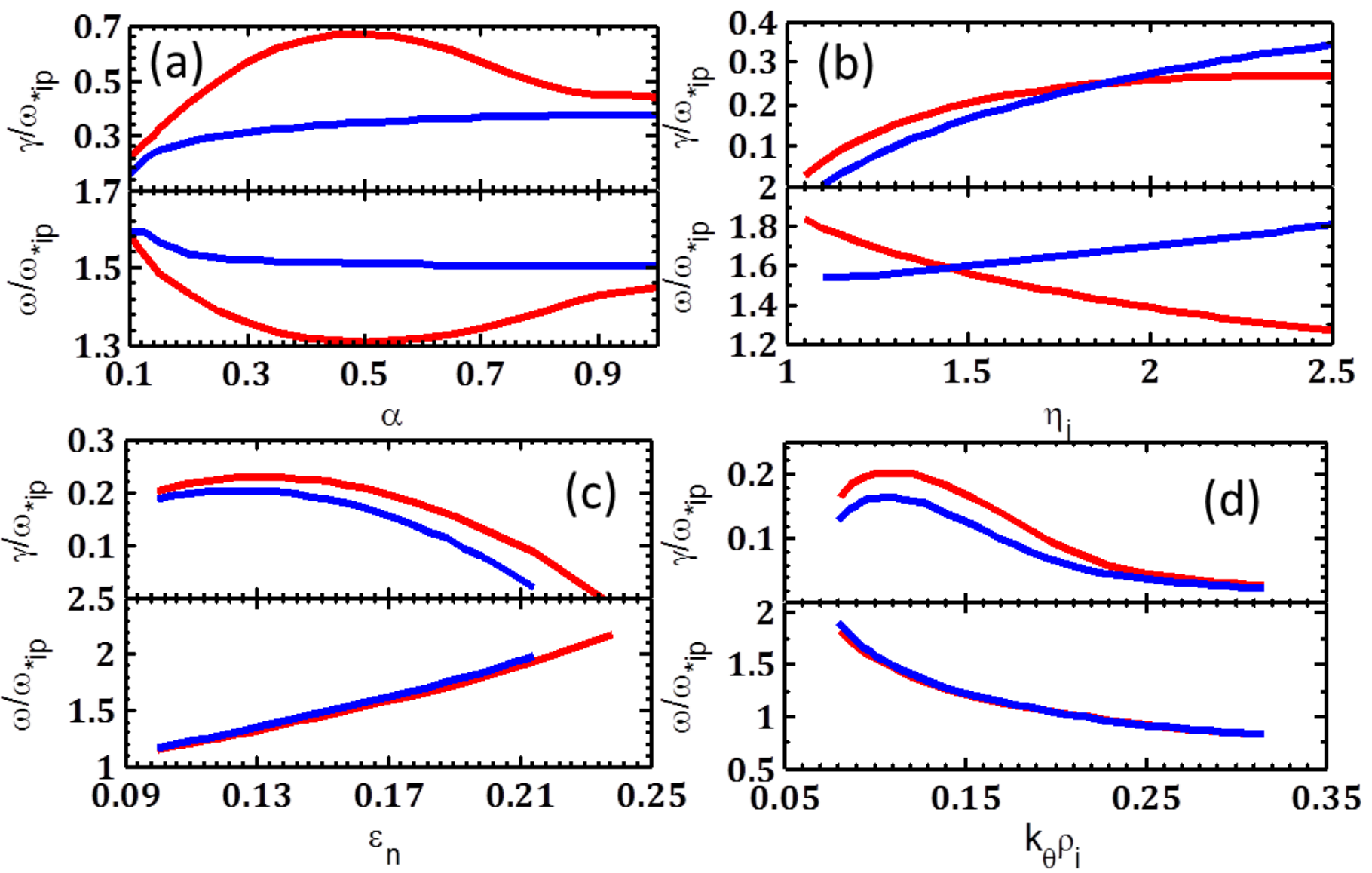}
\caption{\label{fig1} Real frequecy ($\omega/\omega_{*ip}$) and growth rate ($\gamma/\omega_{*ip}$) vs $\alpha$ (a), $\eta_i$ (b), $\varepsilon_n$ (c) and $k_{\theta}\rho_i$ (d). Parameters all take from $\alpha=0.1$, $\eta_i=1.5$, $-R_0 \bigtriangledown ln n_i\equiv 1/\varepsilon_n =6.0$, $q=1$, and $k_{\theta}\rho_i=0.1$ expect for the scanning parameter in the each subgraph. Red line, $s=0.1$; Blue line, $s=-0.1$.}
\end{figure}

Figure 4 shows solutions of the GFLDR-E according to experimental parameters. These modes are more unstable in the case of large $\eta_i$, $-R_0 \bigtriangledown ln n_i$ and ion diamagnetic effect in the case $\Omega_{*pi}/\sqrt{7/4+\tau}q \simeq 0.83$. At $T_i=0.6keV$, $\omega_r/\omega_{*pi}\simeq 1.3-1.5$ and $\gamma/\omega_{*pi}\simeq 0.1-0.4$. Figure 5 illustrates the diamagnetic effect from the density gradient on the stabity of coherent modes. The mode excitation needs an $\Omega_{*ni}$ threshold, while the mode is stabilized at higher $\Omega_{*ni}$ values. Table I gives a comparison between measured frequency and real frequency predicted by the GFLDR-E, showing excellent agreement. All properties of the observed coherent modes are consistent with their interpretation as AITG modes. To show this and better understand the excitation mechanism of the modes with weak magnetic shear and low pressure gradients, we solve the AITG/KBM equation, i.e.,eq.(2). Figure 6 shows the mode frequency and growth rate at different pressure gradients, $\eta_i$, density gradients and Larmor radius effects. Obviously, for the mode excitation with weak magnetic shear $\eta_i$ and $-R_0 \bigtriangledown ln n_i$ both have thresholds which are responsible for the onset of turbulence and the profile stiffness, and the modes are more unstable in the case of large $\eta_i$ and $-R_0 \bigtriangledown ln n_i$. For weak magnetic shears, the $\alpha$ threshold for mode excitation is very low. Meanwhile, larger $k_{\theta}\rho_i$ values have a stabilizing effect. Figure 7 presents the mode frequency and growth rate at different magnetic shears, and it is found that $s$ has a very narrow window for the unstable modes at $\alpha=0.1$. This window becomes wider for increasing $\alpha$ and the magnetic shear corresponding to the maximum of growth rate shifts towards the positive shear region. Figure 7(b) gives the eigenfunctions of the unstable modes at weak magnetic shear and low pressure gradient. The parallel mode structure is very extended along the magnetic field, which is different with respect to the electrostatic ITG ballooning structure. These analysis results illuminate why we observed coherent modes only in plasmas with peaked density profiles and with $|s|\ll1$.

\begin{figure}[!htbp]
\centering
\includegraphics[scale=0.8]{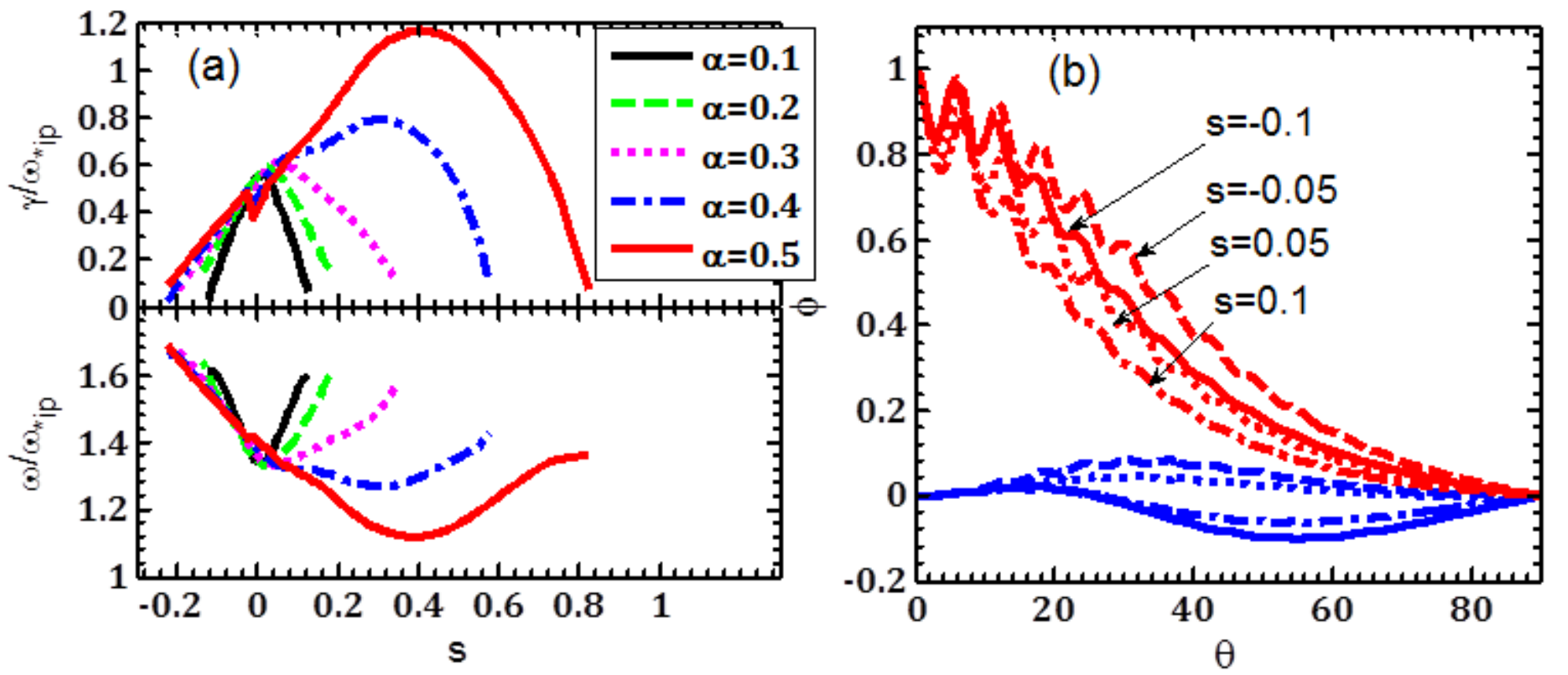}
\caption{\label{fig1} Real frequecy ($\omega/\omega_{*ip}$) and growth rate ($\gamma/\omega_{*ip}$) of AITG modes vs $s$ at the different $\alpha$ (a); eigenfunctions of AITG modes at the different $s$ and $\alpha=0.1$ (red line, $Re(\phi)$; blue line, $Im(\phi)$)(b). Other parameters: $\eta_i=1.5$, $-R_0 \bigtriangledown ln n_i =6.0$, $q=1$, and $k_{\theta}\rho_i=0.1$.}
\end{figure}

The AITG modes can become unstable for $\eta_i$ larger than a critical value $\eta_{ic}$ which is given by $\eta_{ic}\simeq 2/\sqrt{7+4\tau}q \Omega_{*ni}$\cite{zcsppcf96}. For our experiments, $\Omega_{*pi}< \sqrt{7/4+\tau}q$ is satisfied. With the density-peaking and $T_i$-decreasing at the q=1 surfaces, $\eta_{ic}$ drops. Therefore the threshold condition weakens and the modes become unstable more easily. For the modes with different mode-numbers, $\eta_{ic}$ may be different so that the frequency staircase occurs during the density ramp-up corresponding to $\eta_{ic}$ sequential decreasing at the marginal stability. With a milder density rump-up and/or constant density, multiple branches of modes may simultaneously appear.

The AITG modes may have important implications on plasma transport and effectively limit the maximum achievable density and pressure in tokamaks. Figure 1 shows that the low frequency kink-tearing mode with $m/n=-2/-1$ or $-1/-1$ grows rapidly after the AITG modes are driven unstable by density peaking. Subsequently, the bulk plasma produces a minor disruption. This suggests that the disruption is potentially linked with the nonlinear evolution of these instabilities, however the corresponding physical mechanism is unclear at present.

In summary, we report experimental observation of AITG instabilities in HL-2A Ohmic plasmas. A group of coherent modes with $f=15-40 kHz$ and $n=3-6$ is consistently measured by multiple diagnostics in high-density Ohmic regimes. They arise in plasmas with peaked density and weak magnetic shear. The instabilities are excited by pressure gradients. Different unstable modes can be excited at different times during density ramp-up, when plasma conditions pass through marginal stability, yielding the characteristic signature of a frequency staircase. Meanwhile, at nearly constant plasma density and radial profiles, multiple modes coexist and are simultaneously observed. Theoretical analyses by the GFLDR-E reveal that mode frequencies scale with ion diamagnetic drift frequency and $\eta_i$, and they lie in the KBM-AITG-BAE frequency range. These AITG modes are more unstable when the magnetic shear is small in low pressure gradient regions. AITG/KBM equation also illuminate why AITG modes can be unstable for weak shear and low pressure gradients. The low-$n$ AITG modes are thermal plasma ion wave-particle interaction mediated by geodesic curvature coupling and, thus, observed in experiments due to weak magnetic shear and low pressure gradient. With increasing $s$ and $\alpha$, the toroidal mode-number of the most unstable mode also increases, and coherent modes gradually evolve into Alfv{\'e}nic turbulence. The AITG is an electrostatic ITG counterpart, and, within an unstable window, its growth rate is larger than that of the ITG mode which is clearly stabilized by finite-$\beta$ effects. The threshold for AITG destabilization is typically lower than that of ideal marginal stability($\alpha_{AITG}/\alpha_c\simeq 0.5$)\cite{zcsdppcf98}. The stability of AITG modes has a strong and complex dependence on values of $s$ and $\alpha$, and various effects including trapped electron, finite $B_{\parallel}$, plasma shape, Shafranov shift and parallel ion current\cite{hmpop,fvvpop03,gavvpop04,moralinf14}. Thus, a full and thorough assessment of the mode stability requires a kinetic global simulation using the realistic configuration and profiles. The synergetic effect of $s$ and $\alpha$ is a dominant factor for the mode stability, and it maybe play an important role in the formation and evolution of internal/external transport barriers (ITB /ETB, corresponding to large $\eta_i$ and strong $-R_0 \bigtriangledown ln n_e$ respectively) with weak and negative magnetic shears. Similar AITG/KBM phenomena have been observed in DIII-D QH-mode plasmas\cite{yanzprl11} and HL-2A ITB plasmas with weak magnetic shear. However, observations reported in the present work are the first clear experimental identification of this phenomenology, fully consistent with theoretical interpretation and numerical stability analyses. In addition, it needs to be stressed that these instabilities maybe also exist in the saturated Ohimic confinement (SOC) regime and density limit plasmas. The interaction between AITG/KBM activities and EPs should also be investigated with greater attention in fusion plasmas, such as ITER, since weak magnetic shear amplifies the role of and possible excitation by EP of these fluctuations, as predicted by the GFLDR-E. This work is the first clear experimental evidence of AITG/KBM  and complex plasma behaviors fully consistent with the theoretical framework of the GFLDR-E. It also paves the road to more in depth analyses of similar phenomena in fusion plasmas with non-perturbative EP populations, with suggestive possibility of controlling plasma performance by a careful choice of plasma profiles in the weak shear core region typical of burning fusion plasmas.

\vspace{2ex}
One of the authors (C.W.) are very grateful to the HL-2A group. This work is supported in part by the ITER-CN under Grants No. 2013GB104001 and 2013GB106004, and by NNSF of China under Grants No. 11475058.

\end{spacing}


\begin{thebibliography}{99}

  \bibitem{gvlc}
G. Vetoulis and L. Chen,
Journal of Geophysical Research: Space Physics {\bf 101}, 15441  (1996).

  \bibitem{czrmp}
  L. Chen and F. Zonca,
Rev. Mod. Phys. {\bf 88}, 015008  (2016).

  \bibitem{wu12}
  D. Wu, Kinetic Alfv{\'e}n Wave: Theory, Experiment and Application,
  Scientific Press, Beijing, (2012).

  \bibitem{lauberpr}
P. Lauber,
Phys. Rep. {\bf 533}, 33  (2013).

 \bibitem{nsk}
N. N. Gorelenkov, S. D. Pinches and K. Toi,
Nucl. Fusion {\bf 54}, 125001  (2014).

\bibitem{zoncapop99}
F. Zonca {\sl et al.},
Phys. Plasmas {\bf 6}, 1917  (1999).

\bibitem{dongpop99}
J. Q. Dong, L. Chen and F, Zonca,
Nucl. Fusion {\bf 39}, 1041  (1999).

\bibitem{chengczpof}
C. Z. Cheng,
Phys. Fluids {\bf 25}, 1020  (1982).

\bibitem{hmprl}
A. Hirose and M. Elia,
Phys. Rev. Lett. {\bf 76}, 628  (1996).

\bibitem{hmpop}
A. Hirose and M. Elia,
Phys. Plasmas {\bf 10}, 1195  (2003).

\bibitem{gvprl}
R. Ganesh and J. Vaclavik,
Phys. Rev. Lett. {\bf 94}, 145002  (2005).

\bibitem{liyg}
Y. G. Li {\sl et al.},
Plasma Sci. Technol. {\bf 17}, 430  (2015).

\bibitem{jiangm}
 M. Jiang {\sl et al.},
 Rev. Sci. Instrum. {\bf 84}, 113501 (2013).

\bibitem{shipw}
P. W. Shi {\sl et al.},
Plasma Sci. Technol. {\bf 18}, 708  (2016).

\bibitem{cht78}
J. W. Connor and R. J. Hastie and J. B. Taylor,
Phys. Rev. Lett. {\bf 40}, 396  (1978).

  \bibitem{zcsppcf96}
F. Zonca, L. Chen and R. A. Santoro,
Plasma Phys. Control. Fusion {\bf 38}, 2011  (1996).

  \bibitem{zcpop14}
F. Zonca and L. Chen
Phys. Plasmas {\bf 21}, 072120 and 072121 (2014).

  \bibitem{mzcpop15}
R. R. Ma. F. Zonca and L. Chen,
Phys. Plasmas {\bf 22}, 092501 (2015).

\bibitem{zcsdppcf98}
F. Zonca {\sl et al.},
Plasma Phys. Control. Fusion {\bf 38}, 2009  (1998).

\bibitem{fvvpop03}
G. L. Falchetto, J. Vaclavik and L. Villard,
Phys. Plasmas {\bf 10}, 1424 (2003).

\bibitem{gavvpop04}
R. Ganesh {\sl et al.},
Phys. Plasmas {\bf 11}, 3106 (2004).

 \bibitem{moralinf14}
S. Moradi {\sl et al.},
Nucl. Fusion {\bf 54}, 123016  (2014).

\bibitem{yanzprl11}
Z. Yan {\sl et al.},
Phys. Rev. Lett. {\bf 107}, 055004  (2011).





\end{thebibliography}
\end{document}